\documentclass[aps,prd,12pt,notitlepage,showpacs,nofootinbib,tightenlines]{revtex4-1}
\usepackage{amsmath}
\usepackage{amssymb}
\usepackage{epsfig}
\usepackage{graphicx}
\usepackage{bm}
\usepackage{times}
\usepackage{braket}
\usepackage{color}
\usepackage{slashed}
\usepackage{hyperref}
\definecolor{Red}{rgb}{1.,0.,0.}

\definecolor{Blue}{rgb}{0.,0.,1.}

\definecolor{nicered}{rgb}{0.7,0.1,0.1}
\definecolor{nicegreen}{rgb}{0.1,0.5,0.1}
\hypersetup{colorlinks,citecolor=nicegreen,linkcolor=nicered}

\begin{document}
\newcommand{\beq}{\begin{eqnarray}}
\newcommand{\eeq}{\end{eqnarray}}
\newcommand{\ben}{\begin{enumerate}}
\newcommand{\een}{\end{enumerate}}
\newcommand{\be}{\begin{equation}}
\newcommand{\ee}{\end{equation}}
\newcommand{\non}{\nonumber\\ }

\newcommand{\jpsi}{J/\Psi}
\newcommand{\ppa}{\phi_\pi^{\rm A}}
\newcommand{\ppp}{\phi_\pi^{\rm P}}
\newcommand{\ppt}{\phi_\pi^{\rm T}}
\newcommand{\ov}{ \overline }

\newcommand{\zerot}{ {\textbf 0_{\rm T}} }
\newcommand{\kt}{k_{\rm T} }
\newcommand{\fb}{f_{\rm B} }
\newcommand{\fk}{f_{\rm K} }
\newcommand{\rk}{r_{\rm K} }
\newcommand{\mb}{m_{\rm B} }
\newcommand{\mw}{m_{\rm W} }
\newcommand{\im}{{\rm Im} }

\newcommand{\kks}{K^{(*)}}
\newcommand{\acp}{{\cal A}_{\rm CP}}
\newcommand{\pb}{\phi_{\rm B}}

\newcommand{\xeba}{\bar{x}_2}
\newcommand{\xsba}{\bar{x}_3}
\newcommand{\peas}{\phi^A}

\newcommand{\pvsl}{ p \hspace{-2.0truemm}/_{K^*} }
\newcommand{\esl}{ \epsilon \hspace{-2.1truemm}/ }
\newcommand{\psl}{ p \hspace{-2truemm}/ }
\newcommand{\ksl}{ k \hspace{-2.2truemm}/ }
\newcommand{\lsl}{ l \hspace{-2.2truemm}/ }
\newcommand{\nsl}{ n \hspace{-2.2truemm}/ }
\newcommand{\vsl}{ v \hspace{-2.2truemm}/ }
\newcommand{\epsl}{\epsilon \hspace{-1.8truemm}/\,  }
\newcommand{\bfkk}{{\bf k} }
\newcommand{\calm}{ {\cal M} }
\newcommand{\calh}{ {\cal H} }
\newcommand{\calo}{ {\cal O} }

\def \appb{{\bf Acta. Phys. Polon. B }  }
\def \cpc{ {\bf Chin. Phys. C } }
\def \ctp{ {\bf Commun. Theor. Phys. } }
\def \epjc{{\bf Eur. Phys. J. C} }
\def \ijmpcs{{\bf Int. J. Mod. Phys. Conf. Ser.} }
\def \jhep{{\bf J. High Energy Phys. } }
\def \jpg{ {\bf J. Phys. G} }
\def \mpla{{\bf Mod. Phys. Lett. A } }
\def \npb{ {\bf Nucl. Phys. B} }
\def \plb{ {\bf Phys. Lett. B} }
\def \ppn{ {\bf Phys. Part. Nucl. } }
\def \ppnp{{\bf Prog.Part. Nucl. Phys.  } }
\def \pr{  {\bf Phys. Rep.} }
\def \prc{ {\bf Phys. Rev. C }}
\def \prd{ {\bf Phys. Rev. D} }
\def \prl{ {\bf Phys. Rev. Lett.}  }
\def \ptp{ {\bf Prog. Theor. Phys. }}
\def \zpc{ {\bf Z. Phys. C}  }
\def \jpg{ {\bf J.Phys.-G-}  }
\def \ap{ {\bf Ann. of Phys}  }
\def \pr{ {\bf Phys. Rev.}  }
\def \jmp{ {\bf J. Math. Phys.}  }
\def \ap{{\bf Ann. Phys.}}
\title{The next-to-leading order corrections to $\rho$ meson electromagnetic form factors in the $k_T$ factorization
approach}
\author{Ya-Lan Zhang$^{1}$} \email{zylyw@hyit.edu.cn}
\author{Jun Hua$^{2}$} \email{546406604@qq.com}
\author{Dong-Sheng Li$^{1}$}
\author{Zhen-Jun Xiao$^{2}$ } \email{xiaozhenjun@njnu.edu.cn}
\affiliation{1. Department of Faculty of Mathematics and Physics, Huaiyin Institute of Technology,
Huaian,Jiangsu 223200, People's Republic of China,}
\affiliation{2.  Department of Physics and Institute of Theoretical Physics,
Nanjing Normal University, Nanjing, Jiangsu 210023, People's Republic of China,}
\date{\today}
\begin{abstract}
In this paper we calculate the next-to-leading-order (NLO) corrections to $\rho$-meson electromagnetic form factors
by employing the $k_T$ factorization approach.
We find that the NLO correction to $F_i (Q^2)(i=LT,TL)$  is around $30\%$ of the leading-order (LO) contribution
in the region $Q^2>2GeV^2$. The NLO correction to  $F_{LL}(Q^2)$ is close to $20\%$ of the LO one  in the region
$Q^2>3GeV^2$.
The NLO radiative corrections to  the electric, magnetic, and quadruple form factors $F_j(Q^2) (j=1,2,3)$ are
sizeable in magnitude and agree with those from other approaches.
\end{abstract}



\maketitle

\section{Introduction}

The form factors, as the important nonperturbative observables in hadron physics, are essential for studying the structure of various hadrons.
For light meson, their electromagnetic and transition form factors have been studied extensively due to their significance in phenomenology.
Over the past decades, the collinear and $k_T$ factorization for light meson form factors have been demonstrated
\cite{pi2,Nagashima2,Nagashima3,rhopi,rho,Brho} and calculated\cite{pipi2,pipi3,rholo,rhopinlo} based on the Perturbative QCD (PQCD) factorization
approach. For the pseudoscalar mesons such as the pion, the form factors  are well understood up to NLO accuracy \cite{pipi2,pipi3}.
For the vector mesons, however, the situation is rather different.
For the $\rho $ meson, for example, the NLO correction to its electromagnetic form factor in $k_T$ factorization is still absent now.
In early years, the NLO correction to $\rho$-meson electromagnetic form factors have been investigated by employing the QCD sum rules\cite{Braguta}, and the light cone sum rules\cite{Aliev},the light-front quark model \cite{Ho-Meoyng}, lattice calculate \cite{W. Andersen} and so on.
In this paper, we will calculate the NLO correction to the $\rho$-meson electromagnetic form factors associated with
the process $\rho\gamma^* \to \rho$ in the framework of the $k_T$ factorization approach.

A modified perturbative QCD approach based on the $k_T$ factorization, named as the PQCD approach
\cite{Brodsky,Huang,Li381,Botts,StermanAJ,YuHP,ZuoHZ}, was proposed two decades ago and has been successfully applied for many exclusive
processes, such as the calculations of some transition form factors \cite{Zhang2014}, the two body and three body decays
of B meson \cite{wang2016,Rui2018,Ya2017,Ma2017,Qin:2014xta}, the production of light mesons in  colliders \cite{Lu:2018obb} and so on.
By introducing a small transverse momentum $k_T$, the infrared divergences could be regulated in both the full QCD Feynman diagrams and the effective diagrams.
Both the large double logarithms could be absorbed by the resummation technology.
The light-cone singularities are regularized by rotating the Wilson lines away from the light cone \cite{collins, J.P. Ma}.
The physical amplitudes then can be written as the convolution of the universal nonperturbative hadron wave functions and a hard scattering kernel.
For the $\rho$ meson electromagnetic form factor, its NLO hard kernel is defined as the difference
between the full QCD Feynman diagrams and the corresponding effective ones in $k_T$ factorization theorem.
In the transition process, the collinear divergence from gluon
emission in the quark diagrams are cancelled by those divergences in the $\rho$ effective wave function, and the soft divergences
will be canceled by summing over all quark diagrams, which ensured by the KLN theorem
\cite{Lee:1964is,Kinoshita:1962ur}.
Then we can derive the infrared-safe $k_T$-dependent NLO hard kernel for $\rho$-meson
electromagnetic form factors in the $k_T$ factorization approach, which setting the renormalization and factorization scales as the internal hard scale.

This paper is organized as follows. In Sec.~II, we calculate the $O(\alpha_s^2)$ QCD quark diagrams, and effective diagrams for rho wave function, and then derive the NLO hard kernel for its electromagnetic form factor.
Numerical results are performed in Sec.~III.
Sec.~IV contains the conclusion.

\section{THEORETICAL FRAMEWORK }

\begin{figure}[htbp]
\begin{center}
\vspace{-1.5cm}
\includegraphics[width=0.8\textwidth]{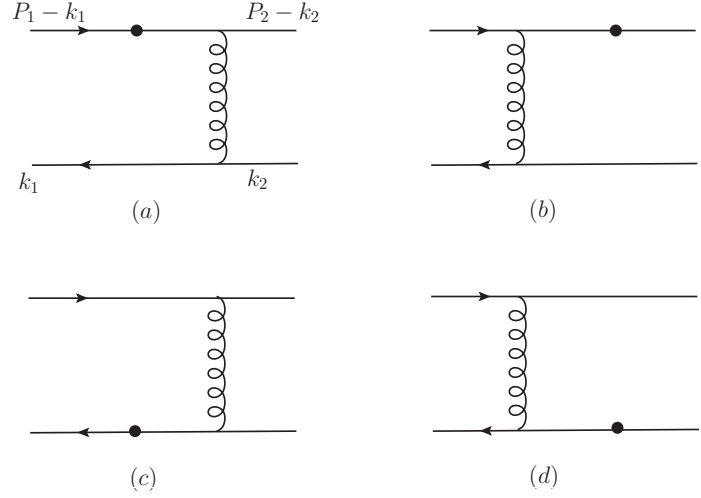}
\vspace{-10cm}
\caption{The leading order Feynman diagrams for $\rho\gamma^{\star}\rightarrow\rho$ process.}
\label{fig:fig1}
\end{center}
\end{figure}

In this section we consider the NLO gluon radiative corrections to the rho meson electromagnetic form factors in the framework of the
$k_T$ factorization.
The moment of initial-state (final-state) $\rho$ meason is defined in the light-cone (LC) coordinate as,
\begin{align}
p_1=\frac{Q}{\sqrt{2}}(1,r^2_{\rho},0),\,\,\,\,\,\
p_2=\frac{Q}{\sqrt{2}}(r^2_{\rho},1,0),
\end{align}
where $Q^2=2P_1\cdot P_2$ is the momentum transfer squared from the virtual photon, and the dimensionless factor $r^2_\rho$ 
is $r^2_{\rho}\equiv M^2_{\rho}/Q^2 $.
The anti-quark in initial and final pion carries momentum $k_1=(x_1p^+_1,0,\mathbf{k_{1T}})$ and $k_2=(0,x_2P^-_2,\mathbf{k_{2T}})$, respectively, $\mathbf{k_T}$
represents the transversal momentum, and x is parton momentum fraction.
The polarization vectors of the initial and final $\rho$ meson are also defined as,
\begin{align}
\epsilon_{1 \mu}(L) = \frac{1}{\sqrt{2} r_{\rho}}(1,-r_{\rho}^2,\mathbf{0_{T}}),\,\,\,\,\
\epsilon_{1 \mu}(T) =  (0,0,\mathbf{1_{T}}),\non
\epsilon_{2 \mu}(L) = \frac{1}{\sqrt{2} r_{\rho}}(-r_{\rho}^2,1,\mathbf{0_{T}}),\,\,\,\,\
\epsilon_{2 \mu}(T) =  (0,0,\mathbf{1_{T}}).
\end{align}
The LO quark diagrams are shown in Fig.~\ref{fig:fig1}.
We focus on Fig.~\ref{fig:fig1}(a) only, since the contribution from other three diagrams can be obtained  from the exchanging symmetry
between Fig.~\ref{fig:fig1}(a) and Figs.~\ref{fig:fig1}(b-d).

The related meson wave functions are written as in Refs.~\cite{Ball:1998sk,Ball:1998ff},
\beq
&&\langle 0 \vert \bar{u}(0)_j d(z_1)_l \vert \rho^-(p_1,\epsilon_{1T}) \rangle = \frac{1}{\sqrt{2N_C}} \int_0^1 dx_1 e^{ix_1p_1z_1}
\left\{\psl_1 \esl_{1T} \phi_\rho^T(x_1) + m_\rho \esl_{1T} \phi^v_\rho(x_1) \right.  \non
&&\left. \hspace{8cm} + m_\rho i \epsilon_{\mu\nu\rho\sigma} \gamma^\mu \gamma_5 \epsilon^\nu_{1T} n^\rho v^\sigma \phi_\rho^a(x_1)  \right\}_{lj},
\label{eq:1}\\
&&\langle 0 \vert \bar{u}(0)_j d(z_1)_l \vert \rho^-(p_1,\epsilon_{1L}) \rangle = \frac{1}{\sqrt{2N_C}} \int_0^1 dx_1 e^{ix_1p_1z_1}
\left\{ M_{\rho} \esl_{1L} \phi_\rho(x_1) + \esl_{1L}\psl_1 \phi^t_\rho(x_1) \right.\non
&&\left.\hspace{8cm}+M_{\rho} \phi_\rho^s(x_1)  \right\}_{lj} ,
\label{eq:2}\\
&&\langle \rho^+(p_2,\epsilon_{2T}) \vert \bar{u}(z_2)_j d(0)_l \vert  0 \rangle = \frac{1}{\sqrt{2N_C}} \int_0^1 dx_2 e^{ix_2p_2z_2}
\left\{\esl_{2T} \psl_2  \phi_\rho^T(x_2) + m_\rho \esl_{2T} \phi^v_\rho(x_2) \right.  \non
&&\left. \hspace{8cm} + m_\rho i \epsilon_{\mu\nu\rho\sigma} \gamma_5 \gamma^\mu \epsilon^\nu_{2T} n^\rho v^\sigma \phi_\rho^a(x_2)  \right\}_{lj},
\label{eq:3}\\
&&\langle \rho^+(p_2,\epsilon_{2L}) \vert \bar{u}(z_2)_j d(0)_l \vert  0 \rangle  = \frac{1}{\sqrt{2N_C}} \int_0^1 dx_2 e^{ix_2p_2z_2}
\left\{ M_{\rho} \esl_{2L} \phi_\rho(x_2) + \psl_2\esl_{2L} \phi^t_\rho(x_2)\right. \non
&&\left. \hspace{8cm}+M_{\rho} \phi_\rho^s(x_2)  \right\}_{lj} ,
\label{eq:4}
\eeq
where $\phi_{\rho}$ and $\phi^T_\rho$ denote the twist-2 distribution amplitudes (DAs), $\phi^{s/t}_{\rho}$ and $\phi^{v,a}_\rho$ are twist-3 DAs,
light-like vectors $n=(1,0,\mathbf0_T)$ and $v=(0,1,\mathbf0_T)$, $N_c$ is the number of colors.

In small x region , our calculation is based on the following hierarchy of the energy scales \cite{LiNN},
{\small
\beq
Q^2 \gg x_1Q^2 \thicksim x_2Q^2 \gg x_1x_2Q^2 \gg k^2_{1T} \thicksim k^2_{2T}.
\label{eq:9}
\eeq}

The NLO hard kernel $H^{(1)}$ is defined by taking the difference between the full amplitude and the effective diagrams,
where the wave functions in the latter one absorb all infrared (IR) divergence at a certain order of strong coupling,
{\small
\beq
H^{(1)}(x_1,k_{1T},x_2,k_{2T},Q^2) &=& G^{(1)}(x_1,k_{1T},x_2,k_{2T},Q^2) \non
&-& \int{dx^{'}_1 d^{2} k^{'}_{1T} \, \mathbf{\Phi}_I^{(1)}(x_1,k_{1T};x^{'}_1,k^{'}_{1T})} \, \mathcal{H}^{(0)}(x^{'}_1,k^{'}_{1T},x_2,k_{2T},Q^{2}) \non
&-& \int{dx^{'}_2d^{2}k^{'}_{2T} \, \mathcal{H}^{(0)}(x_1,k_{1T},x^{'}_2,k^{'}_{2T},Q^2) \, \mathbf{\Phi}_F^{(1)}(x_2,k_{2T};x^{'}_2,k^{'}_{2T})}.
\label{eq:10}
\eeq}
where $G^{(1)}$ denotes the NLO quark diagrams associated with Fig.~\ref{fig:fig1}(a),
$\Phi_{I/F}^{(1)}$ presents the $O(\alpha_s)$ effective diagrams for initial (final) quark-level wave function,
and $H^{(0)}$ is the LO hard kernel.

\subsubsection{Full amplitudes at NLO}

\begin{figure}[htbp]
  \centering
  \vspace{-1.2cm}
  \begin{center}
  \epsfxsize=14cm\epsffile{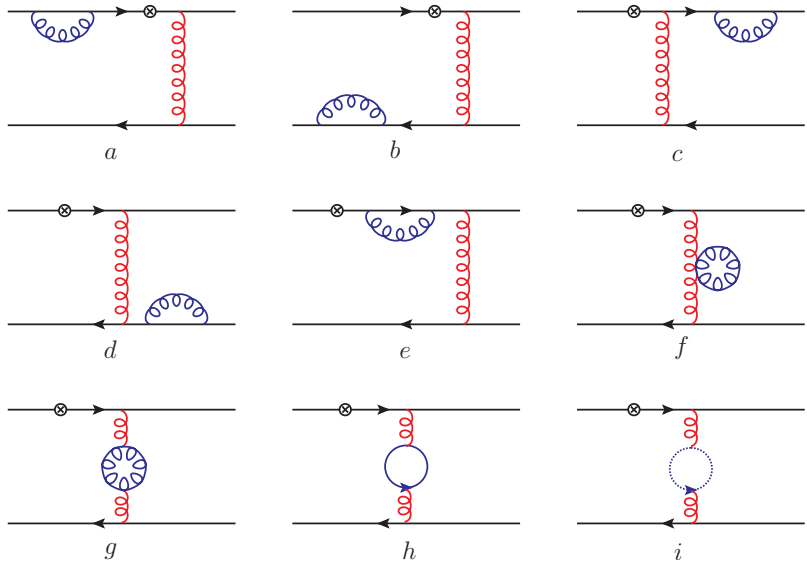}
  \end{center}
  \vspace{-11cm}
\caption{Self-energy corrections to Fig.~\ref{fig:fig1}(a).}
\label{fig:fig2}
\end{figure}

\begin{figure}[htbp]
  \centering
  \vspace{-2cm}
  \begin{center}
  \epsfxsize=14cm\epsffile{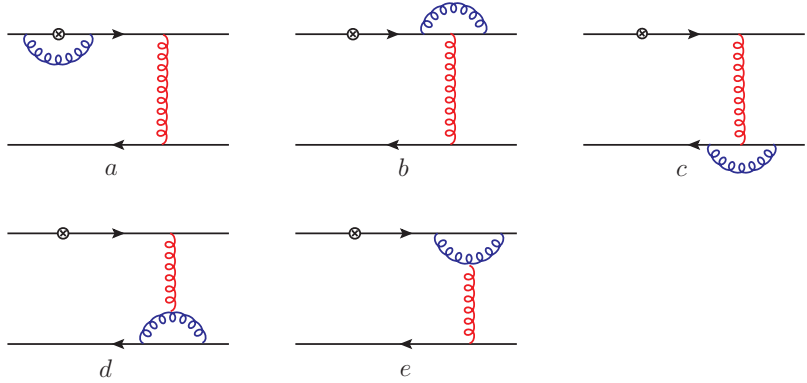}
  \end{center}
  \vspace{-13cm}
\caption{Vertex corrections to Fig.~\ref{fig:fig1}(a).}
\label{fig:fig3}
\end{figure}

\begin{figure}[htbp]
  \centering
  \vspace{-1cm}
  \begin{center}
  \epsfxsize=14cm\epsffile{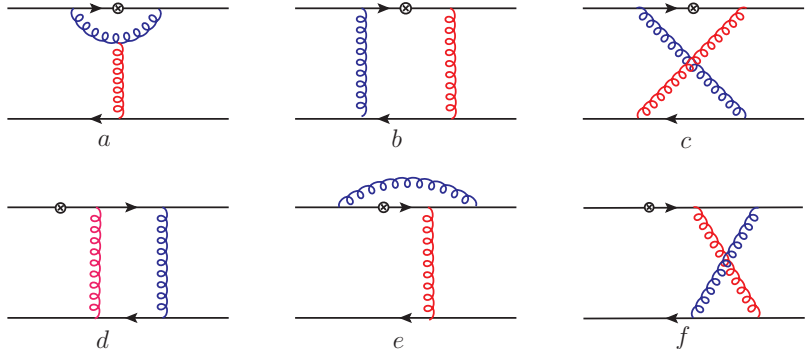}
  \end{center}
  \vspace{-14cm}
\caption{Box and pentagon corrections to Fig.~\ref{fig:fig1}(a).}
\label{fig:fig4}
\end{figure}

The full amplitudes of QCD diagrams for the NLO corrections to Fig.~\ref{fig:fig1}(a) include the self-energy  correction, the vertex correction and the box and pentagon
correction, as shown in Fig.~\ref{fig:fig2}, Fig.~\ref{fig:fig3} and Fig.~\ref{fig:fig4}, respectively.
We firstly define the dimensionless ratios
{\small
\beq
\delta_1=\frac{k^2_{1T}}{Q^2},\quad
\delta_2=\frac{k^2_{2T}}{Q^2},\quad
\delta_{12}=\frac{-(k_{1}-k_{2})^2}{Q^2} \, .
\label{eq:12}
\eeq}
The collinear divergences are both regulated by logarithms $\ln\delta_1$ and $\ln\delta_2$, while
their overlap singularity (the soft divergences) is regulated by double logarithms $\ln^2\delta_1$ and $\ln^2\delta_2$.

We only focus on the NLO correction to $F_{i}(Q^2)$ ($i=LL,TL,LT$), which include the $F_{LL22}$ ( proportional to the term
$\phi_{\rho}(x_1)\phi_{\rho}(x_2)$),
$F_{TL23}$ ( proportional to the term $\phi_{\rho}^T(x_1)\phi_{\rho}^s(x_2)$) and $F_{LT23}$ ( proportional to the term
$\phi_{\rho}(x_1)(\phi_{\rho}^v(x_2)+\phi_{\rho}^a(x_2))$ ).
Fortunately, the structure of the NLO correction to $F_{LL22}$ is the same one as that to pion  electromagnetic form factor
at leading twist \cite{LiNN}.
And the NLO correction to  $F_{TL23}$ has similar behavior with the rho pion transition form factor \cite{rhopinlo}.
So, here we can only pay close attention to the NLO correction to $F_{LT23}$.

By analytic evaluations of the one-loop Feynman diagrams as shown in Fig.~\ref{fig:fig2}(a-i), we find the self-energy corrections of the
external (internal) quark and hard gluon:
{\small
\beq
G^{(1)}_{2a+2b+2c+2d}&=&-\frac{\alpha_s C_F}{4\pi}
\left [\frac{2}{\varepsilon}+\ln\frac{4\pi \mu^2}{\delta_1Q^2 e^{\gamma_E}}+\ln\frac{4\pi \mu^2}{\delta_2Q^2 e^{\gamma_E}}+4 \right]H^{(0)}  \, ,
\label{eq:13} \\
G^{(1)}_{2e}&=&-\frac{\alpha_s C_F}{4\pi}\left [\frac{1}{\varepsilon}+\ln\frac{4\pi \mu^2}{x_1Q^2 e^{\gamma_E}}+2 \right ]H^{(0)} \, ,
\label{eq:14}\\
G^{(1)}_{2f+2g+2h+2i}&=&\frac{\alpha_s C_F}{4\pi}(\frac{5}{3}N_c-\frac{2}{3}N_f)
\left [\frac{1}{\varepsilon}+\ln \frac{4\pi \mu^2}{\delta_{12}Q^2 e^{\gamma_E}}+2\right ]H^{(0)} \, ,
\label{eq:15}
\eeq}
where $1/\varepsilon$ is the ultraviolet pole, $\mu $ is the renormalization scale, $\gamma_E$ is the Euler constant and $N_f$ is the number of
quark flavors.
The collinear divergences from gluon collimated to the initial-state (final-state) external quark in Fig.~\ref{fig:fig2}(a-d) are regularized
into infrared logarithms in $\ln{\delta_1}$ ($\ln{\delta_2}$).
The correction of internal quark is regularized into $ \ln {x_1Q^2} $ in Fig.~\ref{fig:fig2}(e), since it is off-shell by the invariant mass
squared $x_1Q^2$. Analogously, the self-energy correction to the hard gluon is regularized into $\ln\delta_{12}$ in Fig.~\ref{fig:fig2}(f-i).

By the same way, we derived the results for the vertex corrections as shown in Fig.~\ref{fig:fig3}(a-e) :
{\small
\beq
G^{(1)}_{3a}&=&\frac{\alpha_s C_F}{4\pi}\left [\frac{1}{\varepsilon}+\ln\frac{4\pi \mu^2}{Q^2 e^{\gamma_E}}-
(2\ln{\delta_1}+1)\ln{x_1}-\frac{\pi^2}{3}+\frac{3}{2} \right ] H^{(0)} \, ,
\label{eq:16}\\
G^{(1)}_{3b}&=&-\frac{\alpha_s}{8\pi N_c}
\left [\frac{1}{\varepsilon}+\ln\frac{4\pi \mu^2}{x_1 Q^2 e^{\gamma_E}}+3 \right ]H^{(0)} \, ,
\label{eq:17}\\
G^{(1)}_{3c}&=&-\frac{\alpha_s}{8\pi N_c}
\left [\frac{1}{\varepsilon}+\ln\frac{4\pi \mu^2}{\delta_{12} Q^2 e^{\gamma_E}}+3 \right ]H^{(0)} \, ,
\label{eq:18}\\
G^{(1)}_{3d}&=&\frac{\alpha_s N_c}{8\pi}
\left [\frac{3}{\varepsilon}+3\ln\frac{4\pi \mu^2}{\delta_{12} Q^2 e^{\gamma_E}}+
\ln{\frac{\delta_{12}}{\delta_1}}+\ln{\frac{\delta_{12}}{\delta_2}}+3\right ] H^{(0)} \, ,
\label{eq:19}\\
G^{(1)}_{3e}&=&\frac{\alpha_s N_c}{8\pi}\left [\frac{3}{\varepsilon}+3\ln\frac{4\pi \mu^2}{x_{1} Q^2 e^{\gamma_E}}+
\ln{\frac{x_1}{\delta_2}}+2\ln{\frac{x_1}{\delta_{12}}}-2\ln\frac{x_1}{\delta_2}\ln\frac{x_1}{\delta_{12}}-\frac{2}{3} \pi^2+\frac{1}{2} \right ]
H^{(0)} \, .
\label{eq:20}
\eeq}

The NLO correction $G^{(1)}_{3a}$ contains the infrared logarithm $\ln\delta_1$ and $\ln x_1$ only,
since the end of radiative gluon are attached to initial-state external quark line and internal quark line in Fig.~\ref{fig:fig3}(a).
The correction $G^{(1)}_{3b}$ ($G^{(1)}_{3c}$) to the upper (lower) gluon vertex have no infrared logarithms,
because the sequence of $\gamma$-matrices in the fermion flow.
The triple-gluon vertex correction $G^{(1)}_{3d}$ contains the term $\ln (\delta_{12}Q^2)$ with
the LO hard gluon invariant mass squared $\delta_{12}Q^2$.
There are two infrared logarithms $\ln\delta_1$ and $\ln\delta_2$ coming from the gluon collinear with the incoming and outgoing quark lines.
For Fig.~\ref{fig:fig3}(d),  there always exist  a hard gluon at leading order and no overlap of collinear divergence
which may cause the soft divergence (double log).
Another triple-gluon vertex correction $G^{(1)}_{3e}$, the same as $G^{(1)}_{3a}$ in Fig.~\ref{fig:fig3}(a), involves only infrared
logarithm $\ln\delta_2$ and $\ln x_1$.
Unlike the case for Fig.~\ref{fig:fig3}(d),  there is only  infrared logarithm $\ln \delta_2$ in Eq.~(\ref{eq:20}),
which is  mainly induced by the structure of hard gluon restricting  the additional gluon  being attached to the final up quark.
We sum up all the UV terms in the self-energy and vertex corrections into $\alpha_s (11-2N_f/3)/4\pi$, which agree with the
universality of the wave function as given in Ref.~\cite{LiNN}.

The corrections from  the box and pentagon diagrams in Fig.~\ref{fig:fig4} are free from the ultraviolet divergence, and can be written
 in the following forms
{\small
\beq
G^{(1)}_{4a}&=& -\frac{\alpha_s N_c}{8\pi }\left[\ln{\delta_1}-1 \right]H^{(0)}
                -\frac{\alpha_s N_c}{8\pi }\left [\ln{\delta_2}-1 \right ]  \bar{H}^{(0)} \,, \label{eq:22}\\
G^{(1)}_{4c}&=&0\, ,\label{eq:24}\\
G^{(1)}_{4e}&=&\frac{\alpha_s }{4\pi N_c}\left
[\ln{\delta_1}\ln{\delta_2}+\ln{\delta_1}(1-\ln{x_1})-\ln{x_1}+\frac{\pi^2}{6}-1\right] \, , \label{eq:21}\\
G^{(1)}_{4f}&=&-\frac{\alpha_s}{4\pi N_c}\left [\ln{\frac{\delta_{12}}{\delta_1}}\ln{\frac{x_1}{\delta_2}}+\ln{\frac{\delta_{12}}{\delta_2}}+\frac{\pi^2}{6}
\right]H^{(0)} \, ,  \label{eq:23}
\eeq}
where $\bar H^{(0)}$ is the other LO hard kernel, which is induced by the singular gluon attaches to the final-state quark lines.
{\small
\beq
\bar H^{(0)}(x_1,k_{1T},x_2,k_{2T},Q^2) &=&- \frac{16 \pi C_F M^2_{\rho} \alpha_{s}}{3} \frac{\phi_{\rho}(x_1)[\phi^{v}_{\rho}(x_2)+\phi^{a}_{\rho}(x_2)]}{(p_1-k_2)^2(k_1-k_2)^2}.
\label{eq:bh0}
\eeq}
Here we have to point out that the Fig.~\ref{fig:fig4}(a-c) also have the contributions to this LO hard kernel Eq.~(\ref{eq:bh0}).
The correction $G^{(1)}_{4c}$ does not have any infrared logarithms, because the $\gamma$-matrices of the gluon vertex
here would give a suppress by power counting.
The contributions from  the reducible diagrams, Fig.~\ref{fig:fig4} (b) and \ref{fig:fig4}(d), which are  power-suppressed at small x, can be
canceled by the relevant effective diagrams.
The corrections of  Fig.~\ref{fig:fig4}(e) and Fig.~\ref{fig:fig4}(f), with the soft radiative gluon attaching to the incoming and
outgoing quark lines, both involve the soft logarithm $\ln\delta_1\ln\delta_2$, and also cancel with each other.

Then we sum over all the NLO contributions from the LO quark diagrams and find the result
{\small
\beq
G^{(1)}&=& \frac{\alpha_s C_F}{4\pi}\Big [ \frac{1}{\varepsilon}
+\ln\frac{4\pi\mu^2}{Q^2e^{\gamma_E}}-2(\ln\delta_1+\ln\delta_2).-2(\ln x_1\ln\delta_1+\ln x_2\ln\delta_2) \non
&& -\frac{9}{8}\ln x_1-\frac{15}{8}\ln \delta_{12} +\frac{9}{4}\ln x_1\ln x_2+\frac{63}{8}+\frac{3}{16}\pi^2 \Big ] H^{(0)} \, , \label{eq:sum}
\eeq}
for $N_f=6$.

\subsubsection{ Effective diagrams at NLO }

 In this part we  calculate the effective diagrams in terms of the convolution of NLO initial(final) meson wave functions and LO hard kernel,

\begin{figure}[tbp]
  \centering
  \vspace{0cm}
  \leftline{\epsfxsize=12cm\epsffile{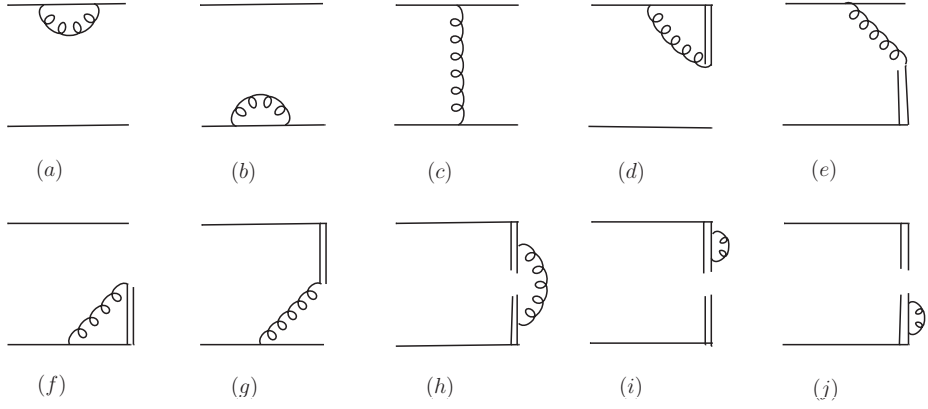}}
  \vspace{-10cm}
\caption{The effective diagrams for the initial $\rho$ meson wave function.}
\label{fig:fig5}
\end{figure}
{\small
\beq
\Phi^{(1)}_\rho \otimes \mathcal{H}^{(0)} &\equiv& \int dx'_1 d^{2} \mathbf{k}'_{1T} \, \Phi^{(1)}_\rho(x_1,\mathbf{k}_{1T};x'_1,\mathbf{k}'_{1T})
\, \mathcal{H}^{(0)}(x'_1,\mathbf{k}'_{1T};x_2,\mathbf{k}_{2T},Q^2),\,\non
\mathcal{H}^{(0)} \otimes \Phi^{(1)}_\rho &\equiv& \int dx'_2 d^{2} \mathbf{k}'_{2T} \,
\mathcal{H}^{(0)}(x_1,\mathbf{k}_{1T};x'_2,\mathbf{k}'_{2T},Q^2)\, \Phi^{(1)}_\rho(x_2,\mathbf{k}_{2T};x'_2,\mathbf{k}'_{2T}),\,
\label{eq:29}
\eeq}\\
and the $O(\alpha_s)$ effective diagrams for the quark-level wave function of $\rho$ meson can be defined as
{\small
\beq
\Phi_{\rho}(x_1,k_{1T};x'_1,k'_{1T})&=&\int \frac{dy^{-}}{2\pi}\frac{d^{2}y_T}{(2\pi)^{2}}
e^{-ix^{'}_1p^{+}_1y^{-}+i \mathbf{k^{'}_{1T}} \cdot \mathbf{y_T}} \non
&& \cdot \langle 0 \vert \bar{q}(y)\gamma_L \vsl W_y^{\dagger}(n_1) I_{n_1;y,0} W_0(n_1) q(0) \vert \bar{u}(p_1-k_1) d(k_1) \rangle \, ,
\label{eq:25}\\
\Phi_{\rho}^{v}(x'_2,k'_{2T};x_2,k_{2T}) &=& \int \frac{dz^{+}}{2\pi}\frac{d^{2}z_T}{(2\pi)^{2}}
e^{-ix^{'}_2p^{-}_2z^{+}+i \mathbf{k^{'}_{2T}} \cdot \mathbf{z_T}} \non
&& \cdot \langle 0 \vert \bar{q}(z)\gamma_T W_z^{\dagger}(n_2) I_{n_2;z,0} W_0(n_2) q(0) \vert \bar{u}(p_2-k_2) d(k_2) \rangle \, ,
\label{eq:26}
\eeq}
where $y=(0,y^-,\mathbf{y_T})$ and $z=(z^+,0,\mathbf{z_T})$ are light cone coordinates of the anti-quark field.
The Wilson line operator $W_y(n_1)$($W_z(n_2))$ with the choice of  $ n_1^2 \ne 0$  ($n_2^2 \ne 0$)
to regularize the light-cone singularities can be written as
{\small
\beq
W_y(n_1) &=& \mathrm{P \, \exp}\left[{-ig_s \int_0^\infty d\lambda n \cdot A(y+\lambda n_1)}\right] \, ,
\label{eq:27}\\
W_z(n_2) &=& \mathrm{P \, \exp}\left[{-ig_s \int_0^\infty d\lambda v \cdot A(z+\lambda n_2)}\right] \, .
\label{eq:28}
\eeq}
Thus the $O(\alpha_s)$ wave function depend on $n^2$ through the scale $\xi_1^2 \equiv 4(n_1 \cdot p_1)^2/\vert n_1^2
\vert$  and $\xi_2^2 \equiv 4(n_2 \cdot p_2)^2/\vert n_2^2 \vert$.
This scale, which should be regarded as a factorization-scheme dependence, can be minimized by choosing a fixed $n^2$\cite{yuming2015}.
In this paper, we will choose the $\xi_1^2=\xi_2^2=Q^2$ to minimize the scheme dependence.

The explicit expressions for the $O(\alpha_s)$ contributions from the effective diagrams displayed in Fig.~\ref{fig:fig5}(a-j) are given as
{\small
\beq
\Phi^{(1)}_{\rho, a} {\otimes} \mathcal{H}^{(0)} &=& \Phi^{(1)}_{\rho,b} \small{\otimes} H^{(0)} =
-\frac{\alpha_s C_F}{8\pi}\left (\frac{1}{\varepsilon}+\ln\frac{4\pi \mu^2_f}{\delta_1 Q^2 e^{\gamma_E}}+2 \right) \, H^{(0)} \,,
\label{eq:30}\\
\Phi^{(1)}_{\rho,c} {\otimes} \mathcal{H}^{(0)} &=& 0 \,,
\label{eq:31}\\
\Phi^{(1)}_{\rho,d} {\otimes} \mathcal{H}^{(0)} &=&\frac{\alpha_s C_F}{4\pi}
\left (\frac{1}{\varepsilon}+\ln\frac{4\pi \mu^2_f}{\xi^2_1 e^{\gamma_E}}
-\ln^2 \frac{k^2_{1T}}{\xi^2_1}-2\ln\frac{k^2_{1T}}{\xi^2_1} +2-\frac{\pi^2}{3} \right ) \, H^{(0)} \,,
\label{eq:32}\\
\Phi^{(1)}_{\rho,e} {\otimes} \mathcal{H}^{(0)} &=&\frac{\alpha_s C_F}{4\pi}
\left (\ln^2 \frac{k^2_{1T}}{x_1 \xi^2_1}+ \pi^2 \right ) \, H^{(0)}  \, ,
\label{eq:33}\\
\Phi^{(1)}_{\rho,f} {\otimes} \mathcal{H}^{(0)} &=&\frac{\alpha_s C_F}{4\pi}\left (\frac{1}{\varepsilon}
+\ln\frac{4\pi \mu^2_f}{\xi^2_1  e^{\gamma_E}}
-\ln^2 \frac{k^2_{1T} }{x^2_1\xi^2_1 }-2\ln\frac{k^2_{1T}}{x^2_1 \xi^2_1} +2-\frac{\pi^2}{3} \right ) \, H^{(0)} \, ,
\label{eq:34}\\
\Phi^{(1)}_{\rho,g} {\otimes} \mathcal{H}^{(0)} &=&\frac{\alpha_s C_F}{4\pi}
\left (\ln^2 \frac{k^2_{1T}}{x^2_1 \xi^2_1}-\frac{\pi^2}{3} \right ) \, H^{(0)}  \,,
\label{eq:35}\\
\Phi^{(1)}_{\rho,h} {\otimes} \mathcal{H}^{(0)} &=&\frac{\alpha_s C_F}{2\pi}
\left (\frac{1}{\varepsilon}+\ln\frac{4\pi \mu^2_f}{\delta_{12} Q^2 e^{\gamma_E}} \right) \, H^{(0)} \, ,
\label{eq:36}
\eeq}
with the factorization scale $\mu_f$.
We can also see that the double log $\ln^2{k_T}$ disappears ultimately due to the same reason as in the full amplitudes.
We naively consider the reducible Fig.~\ref{fig:fig5}(c) as zero because it also reproduces the result of quark diagram Fig.~\ref{fig:fig4}(e) exactly.
Their summation gives
{\small
\beq
\sum_{i=a,\cdots,h} \Phi^{(1)}_{\rho,i} \otimes \mathcal{H}^{(0)}
&=& \frac{\alpha_s C_F}{4\pi} \left[\frac{3}{\varepsilon}
+3\ln\frac{4\pi \mu^2_f}{Q^2e^{\gamma_E}}+\ln\delta_1-2\ln\delta_1+2\ln x_1 \right. \non
&~& \left. \hspace{1cm} +2\ln^2x_1-2\ln\delta_1\ln x_1-2\ln\delta_{12}+2\right] \, H^{(0)}.
\label{eq:37}
\eeq}

We also calculate the third term of Eq.~(\ref{eq:10}), with the wave function of final state meson is Eq.~(\ref{eq:26})
{\small
\beq
\mathcal{H}^{(0)} \otimes \Phi^{(1)}_\rho = \int dx'_2 d^{2}\mathbf{k}'_{2T}  \,
H^{(0)}(x'_1,\mathbf{k}'_{1T};x_2,\mathbf{k}_{2T}) \,
\Phi^{v,a, \,(1)}_\rho(x_2,\mathbf{k}_{2T};x'_2,\mathbf{k}'_{2T}).
\label{eq:38}
\eeq}
The effective Feynman diagrams for the NLO wave function of the final state are similar with those as shown in Fig.~\ref{fig:fig5},
we do not show the details of them for the sake of simplicity, but show the summed result
{\small
\beq
\sum_{i=a,\cdots,h} \mathcal{H}^{(0)} \otimes \Phi^{(1)}_{\rho,i} &=&  \frac{\alpha_s C_F}{4\pi} \left[\frac{3}{\varepsilon}
+3\ln\frac{4\pi \mu^2_f}{Q^2e^{\gamma_E}}+\ln\delta_2-2\ln\delta_2+2\ln x_2 \right. \non
&~& \left. \hspace{1cm} +2\ln^2x_2-2\ln\delta_2\ln x_2-2\ln\delta_{12}+2\right] \, H^{(0)}.
\label{eq:46}
\eeq}

\subsubsection{NLO hard correction}

The double logarithm $\ln^2 x_1$ should be absorbed into the jet function, which emerged when the internal quark is
on-shell in the small $x_1$ region\cite{LiAY},
{\small
\beq
J^{(1)} \, H^{(0)}=-\frac{1}{2} \frac{\alpha_s(\mu_f)C_F}{4\pi} \left[\ln^2x_1+\ln x_1+\frac{\pi^2}{3}\right] \, H^{(0)} \, .
\label{eq:47}
\eeq}
We obtain the NLO hard kernel in the $\overline{\mathrm{MS}}$ scheme with Eq.~(\ref{eq:10}) ,
{\small
\beq
H^{(1)}(\mu, \mu_f, Q^2)  &\rightarrow& H^{(1)} - J^{(1)} \, H^{(0)} \equiv \mathcal{F}_{\rho,LT23}^{(1)}(\mu, \mu_f, Q^2) \, H^{(0)} \non
&=& \frac{\alpha_s(\mu_f) C_F}{4\pi} \left[
\frac{21}{4}\ln{\frac{\mu^2}{Q^2} -6\ln{\frac{\mu_f^2}{Q^2}}} + \frac{9}{4}\ln{x_1}\ln{x_2}  - \frac{21}{8} \ln{x_1} - 2\ln{x_2} \right. \non
&& \left. \hspace{1.7cm} - \frac{1}{2}\ln^2{x_1}- \ln^2{x_2} + \frac{17}{4} \ln{\delta_{12}} + \frac{31}{8} + \frac{17}{48}\pi^2 \right] \, H^{(0)} \,\,\, .
\label{eq:48}
\eeq}
Then we present the expression of $F_{TL23}$ (and $F_{LL22}$) directly, which has the same form with the rho
pion transition form factor ( and pion electromagnetic form factor at leading twist).
{\small
\beq
\mathcal{F}_{\rho,LT32}^{(1)}(\mu, \mu_f, Q^2) \, H^{(0)}
&=& \frac{\alpha_s(\mu_f) C_F}{8\pi} \left[
\frac{21}{2}\ln{\frac{\mu^2}{Q^2} -8\ln{\frac{\mu_f^2}{Q^2}}} + \frac{9}{4}\ln{x_1}\ln{x_2}  - \frac{3}{4} \ln^2{x_1} - \ln^2{x_2} \right. \non
&& \left. \hspace{1.7cm} - \frac{67}{8}\ln{x_1} - 2 \ln{x_2} + \frac{37}{8} \ln{\delta_{12}} + \frac{107}{8} - \frac{\pi^2}{3} \right] \, H^{(0)} \,\,\, ,
\label{eq:50}\\
\mathcal{F}_{\rho,LL22}^{(1)}(\mu, \mu_f, Q^2) \, H^{(0)}
&=& \frac{\alpha_s(\mu_f) C_F}{4\pi} \left[
\frac{21}{4}\ln{\frac{\mu^2}{Q^2} -6\ln{\frac{\mu_f^2}{Q^2}}} + \frac{27}{8}\ln{x_1}\ln{x_2}  - \frac{13}{8} \ln{x_1} + \frac{31}{16}\ln{x_2} \right. \non
&& \left.  \hspace{-2cm}-\ln^2\delta_{12}+\frac{17}{4}\ln x_1\ln\delta_{12}+\frac{23}{8}\ln\delta_{12}
- \frac{17}{4}\ln^2{x_1} + \frac{1}{2} \ln2 + \frac{53}{4} + \frac{5}{48}\pi^2 \right] \, H^{(0)} \,\,\, .
\label{eq:49}
\eeq}

\section{NUMERICAL ANALYSIS}

In this section we evaluate the form factors and present the numerical results.
\beq
F_{\rm LL}(Q^2)&=&\frac{32 \pi C_F}{3} Q^2 \alpha_s(\mu) \int^1_0 dx_1 dx_2 \int^{\infty}_0 b_1 db_1 b_2 db_2
                       ~\cdot \exp[-S_{\rho}(x_i;b_i;Q;\mu)]\non
&~& \times \Bigl \{  \bigl [ x_1-\frac{1}{2}\gamma^2_{\rho}(1+x_1) \bigr ] \phi_{\rho}(x_1)\phi_{\rho}(x_2)[1+\mathcal{F}_{LL}^{(1)}(\mu, \mu_f, Q^2)] + \gamma^2_{\rho}\phi^t_{\rho}(x_1)\phi^s_{\rho}(x_2) \non
&~&                 + 2\gamma^2_{\rho}(1-x_1)\phi^s_{\rho}(x_1)\phi^s_{\rho}(x_2)  \Bigr\} \cdot h(x_2,x_1,b_2,b_1) ,
 \label{eq:rhonff-LL} \\
F_{\rm LT}(Q^2)&=&\frac{32 \pi C_F}{3} Q M_{\rho} \alpha_s(\mu) \int^1_0 dx_1 dx_2 \int^{\infty}_0 b_1 db_1 b_2 db_2
                       ~\cdot \exp[-S_\rho(x_i;b_i;Q;\mu)] \non
&~& \times \Bigl \{  \frac{1}{2}\phi_{\rho}(x_1)\left[\phi^v_{\rho}(x_2)+\phi^a_{\rho}(x_2)\right][1+\mathcal{F}_{LT23}^{(1)}(\mu, \mu_f, Q^2)]\non
                  &&        \ \ \  +\phi^T_{\rho}(x_1)\phi^s_{\rho}(x_2)[1+\mathcal{F}_{LT32}^{(1)}(\mu, \mu_f, Q^2)] \non
&&        \ \ \     - \frac{1}{2}x_1 \phi_{\rho}(x_1) \left[ \phi^v_{\rho}(x_2)+\phi^a_{\rho}(x_2)\right] \Bigr\} \cdot h(x_2,x_1,b_2,b_1),
\label{eq:rhonff-LT}\\
F_{\rm TT}(Q^2)&=&\frac{32 \pi C_F}{3} M^2_{\rho} \alpha_s(\mu) \int^1_0 dx_1 dx_2 \int^{\infty}_0 b_1 db_1 b_2 db_2
                       ~\cdot \exp[-S_\rho(x_i;b_i;Q;\mu)] \non
&~& \times   \Bigl \{  (1-x_1) \left[\phi^a_{\rho}(x_1)\phi^a_{\rho}(x_2)-\phi^v_{\rho}(x_1)\phi^v_{\rho}(x_2)\right ] \non
&~&        \ \ \     + (1+x_1) \left[\phi^a_{\rho}(x_1)\phi^v_{\rho}(x_2)-\phi^v_{\rho}(x_1)\phi^a_{\rho}(x_2) \right ]
\Bigr\} \cdot h(x_2,x_1,b_2,b_1),
\label{eq:rhonff-TT}
\eeq
where the expressions for the the Sudakov factor $S_\rho(x_i;b_i;Q;\mu)$ and the hard function $h(x_2,x_1,b_2,b_1)$ can be found in Ref.~\cite{rholo}.
The NLO correction of $F_{TT}(Q^2)$ is the result of both the initial and final states taking the twist-3 meson distribution amplitudes,
whose power-law behavior is the same as the one combining  twist-2 with twist-4 meson distribution amplitudes \cite{Chengtt}.
So we neglect the  $F_{TT}(Q^2)$  term in this paper.
The light cone distribution amplitudes (LCDAs) are taken up to $n=2$ in the Gegenbauer expansion of the rho meson as given in
Ref.~\cite{Ball},
\beq
\phi_{\rho}(x) &=& \frac{3f_{\rho}}{\sqrt{6}} x(1-x) [1+a^{\|}_{2\rho}C^{3/2}_{2}(t)],\non
\phi^{T}_{\rho}(x)&=& \frac{3f^{T}_{\rho}}{\sqrt{6}} x(1-x) [1+a^{\bot}_{2\rho}C^{3/2}_{2}(t)],\non
\phi^{v}_{\rho}(x)&=&\frac{f_{\rho}}{2\sqrt{6}} [0.75(1+t^{2}) + 0.24(3t^2-1) + 0.12(3-30t^2+35t^4)],\non
\phi^{a}_{\rho}(x)&=&\frac{4f_{\rho}}{4\sqrt{6}} (1-2x) [1+0.93(10x^2-10x+1)];\non
\phi^{t}_{\rho}(x)&=& \frac{f^{T}_{\rho}}{2\sqrt{6}} \left [ 3t^{2} + 0.3t^2(5t^2-3) + 0.21(3-30t^2+35t^4) \right ],\non
\phi^{s}_{\rho}(x)&=& \frac{3f^{T}_{\rho}}{2\sqrt{6}}(1-2x) \left [1 + 0.76\left ( 10x^2-10x+1 \right )\right ],
\label{eq:rho-t3}
\eeq
where $t=2x-1$, the Gegenbauer moments $a_{2,\rho}^\perp = 0.14$, $a_{2,\rho}^\parallel = 0.17$,
and the decay constants $f_\rho = 0.216$, $f_\rho^T=0.160$\cite{StraubICA}. The Gegenbauer polynomials in
Eq.~(\ref{eq:rho-t3}) is $C_2^{3/2}(t)=3/2[5t^2-1]$.
Here, the renormalization and factorization scales are set to the hard scale t which defined in the PQCD
approach to exclusive processes on the $k_{T}$ factorization theorem,
{\small
\beq
\mu=\mu_{f}=t=\max(\sqrt{x_1}Q,\sqrt{x_2}Q,1/b_1,1/b_2).
\label{eq:scale}
\eeq}

\begin{figure}[htbp]
\begin{center}
\vspace{0.4cm}
\includegraphics[width=0.48\textwidth]{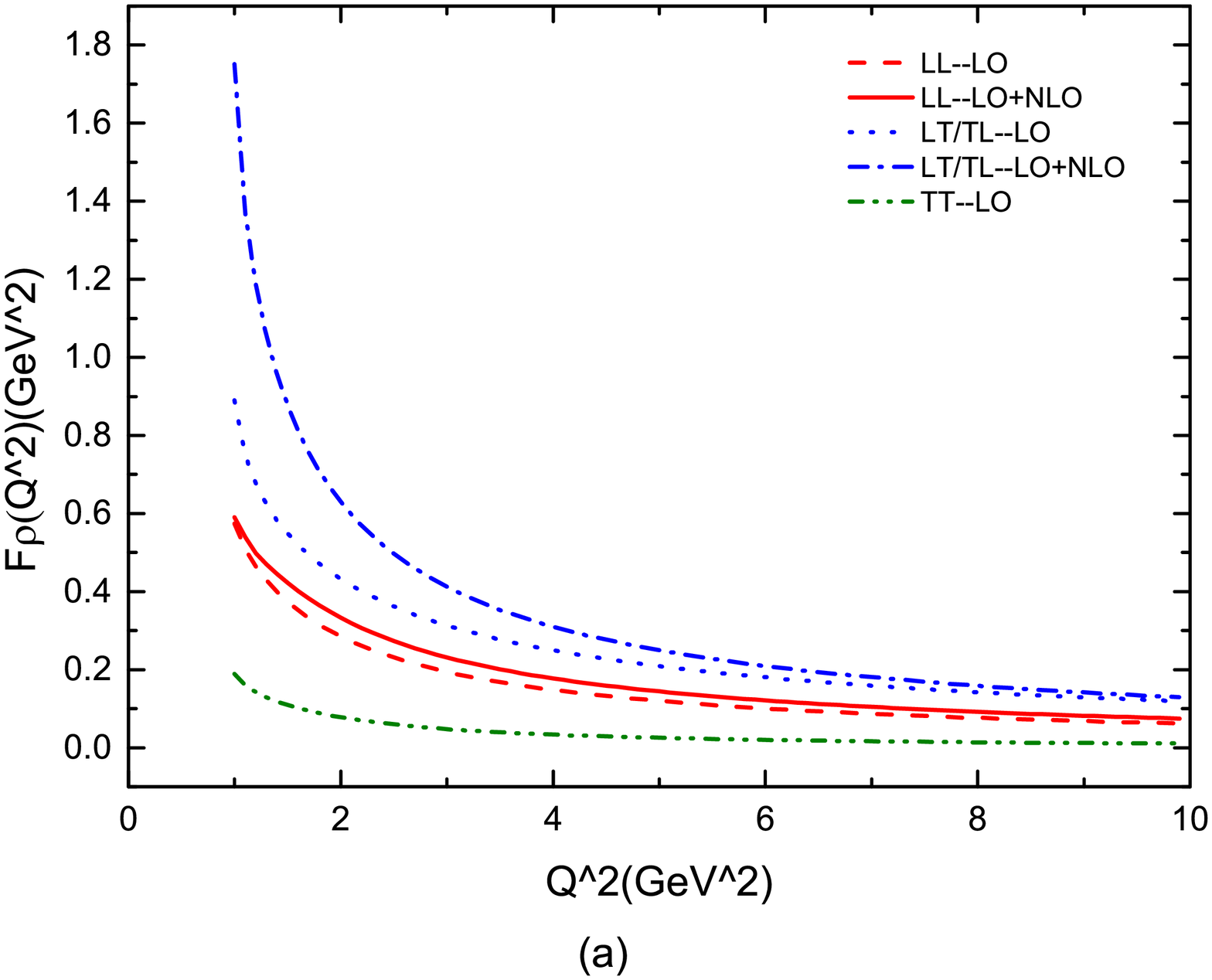}
\hspace{4mm}
\includegraphics[width=0.48\textwidth]{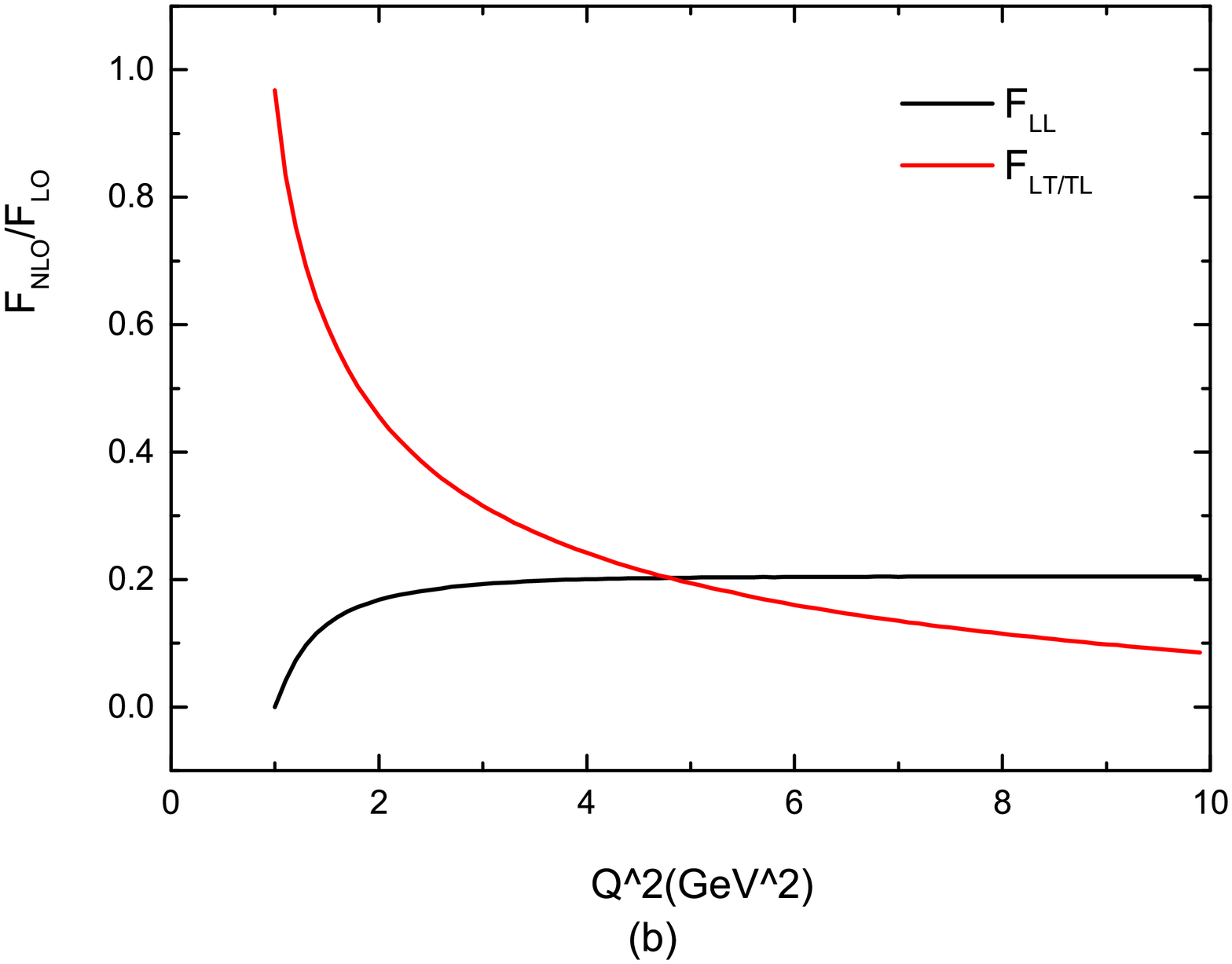}
\hspace{4mm}
\caption{ (a) The PQCD predictions for $Q^2$ dependence of the $\rho$-meson form factors
$F_j(Q^2)(j=LL,LT/TL,TT)$ with different polarizations. (b) The ratio of the NLO contribution over the LO one for the $\rho$ form factors.}
\label{fig:fig6}
\end{center}
\end{figure}

We calculate the LO contributions and NLO contributions in
Eqs.~(\ref{eq:rhonff-LL},\ref{eq:rhonff-LT}), and show the theoretical prediction in Fig.~\ref{fig:fig6}(a):
(1) the red dashed curve denotes the LO contribution of $F_{LL}(Q^2)$, while the red solid  curve shows the summation  of the LO and
NLO contribution; (2) the blue dotted ( dot-dashed ) curve denotes the LO contribution ( the summation of the LO and NLO ones )
of $F_{LT,TL}(Q^2)$; and (3)  finally the green dot-dot-dashed  curve shows the  LO contribution of $F_{TT}(Q^2)$.
We also give the ratio of the NLO contribution over the LO contribution to the $F_{LL}(Q^2)$ and $F_{LT/TL}(Q^2)$
in Fig.~\ref{fig:fig6}(b), where the solid line denotes the ratio of the NLO contribution over the LO contribution to the $F_{LL}(Q^2)$,
and $F_{LT/TL}(Q^2)$.

One can see that the PQCD predictions vary quickly in the  region of $Q^2 < 2GeV^2$,  which tell us that the perturbation theory may not be
reliable in such region.
For $F_{LL}(Q^2)$ the NLO part  is around $ 20\%$ of the LO one when $Q^2 > 2GeV^2$.
For $F_{LT/TL}(Q^2)$, the NLO contribution is around $ 30\% $ of the LO contribution in the region of $Q^2>3GeV^2$.

\begin{figure}[htbp]
\begin{center}
\vspace{0.4cm}
\includegraphics[width=0.49\textwidth]{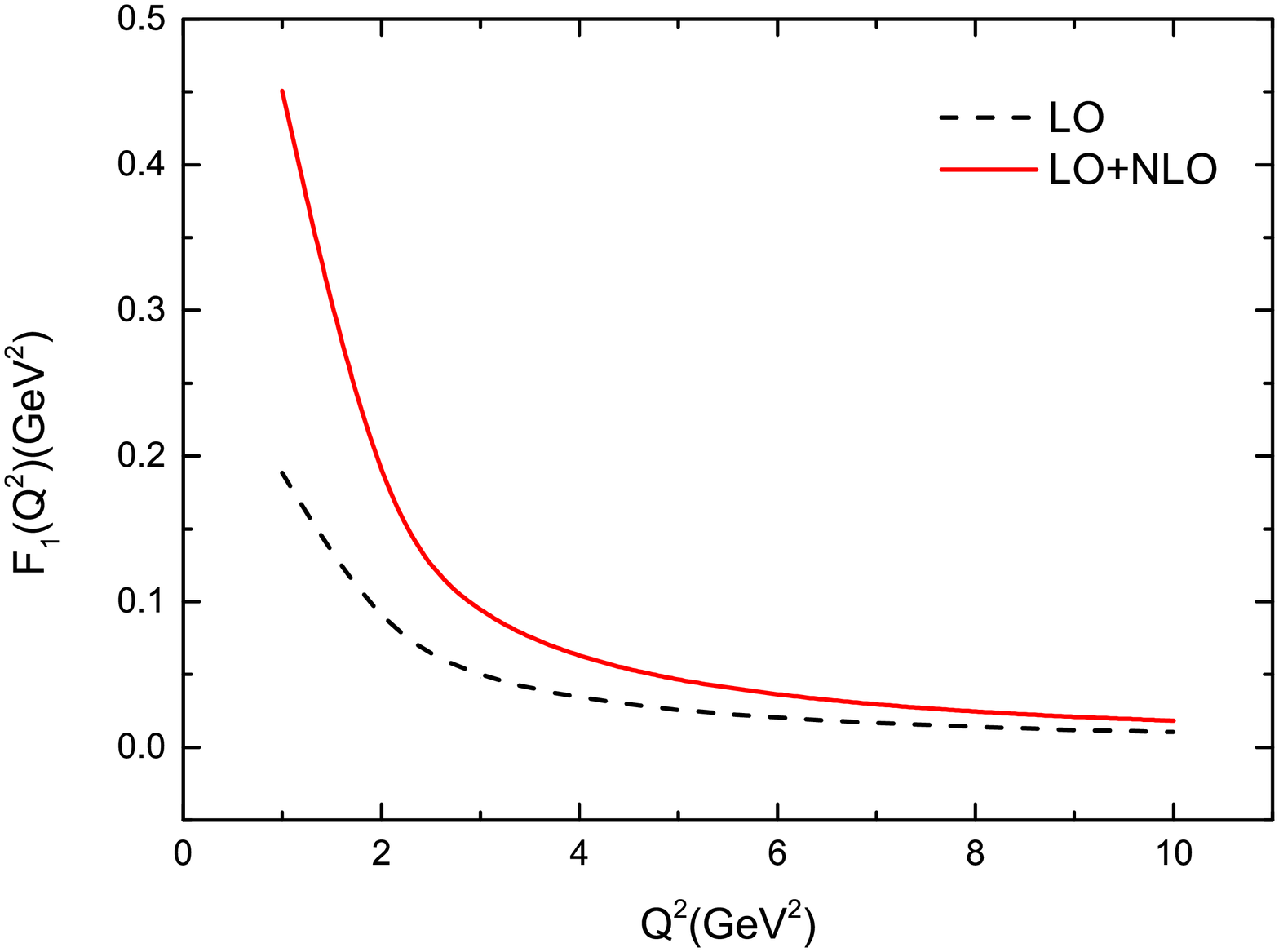}
\caption{(color online). The PQCD predictions for $Q^2$ dependence of electric form factor $F_1$.}
\label{fig:figff1}
\end{center}
\end{figure}

\begin{figure}[htbp]
\begin{center}
\vspace{0.4cm}
\includegraphics[width=0.49\textwidth]{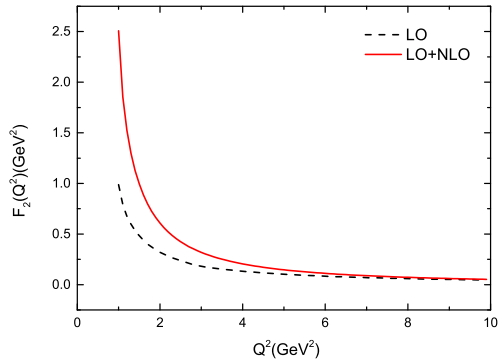}
\hspace{4mm}
\caption{(color online). The PQCD predictions for $Q^2$ dependence of magnetic form factor $F_2$ .}
\label{fig:figff2}
\end{center}
\end{figure}

\begin{figure}[htbp]
\begin{center}
\vspace{0.4cm}
\includegraphics[width=0.49\textwidth]{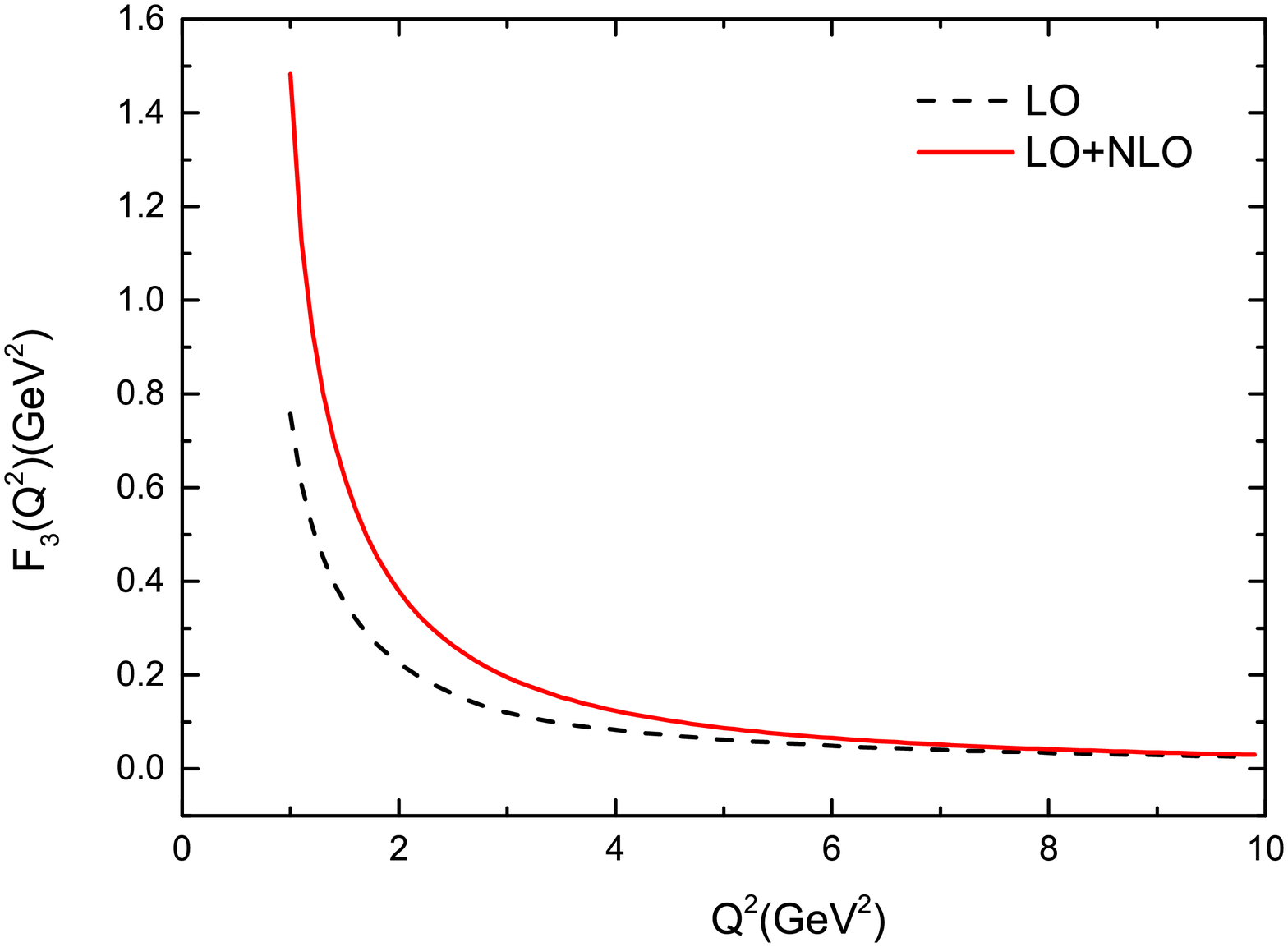}
\vspace{4mm}
\caption{(color online). The PQCD predictions for $Q^2$ dependence of quadrupole form factor $F_3$.}
\label{fig:figff3}
\end{center}
\end{figure}

The results for $\rho$ meson electric $F_1(Q^2)$, magnetic $F_2(Q^2)$ and quadrupole $F_3(Q^2)$  form factors including NLO corrections are shown in Fig.~\ref{fig:figff1}, Fig.~\ref{fig:figff2} and Fig.~\ref{fig:figff3} (solid line denotes LO contribution and curve with long dashes is the sum of LO and NLO contribution).
Apparently at small $Q^2$, our numerical result is not stable.
But the value of our prediction is gradually level off at $Q^2$ greater than $4GeV^2$.
The asymptotic behavior of $\rho$ meson form factors, which derived from quark counting and chirality conservation, are $F_1(Q^2) \sim F_2(Q^2) \sim 1/Q^4$ and $F_3(Q^2) \sim 1/Q^6$.
Obviously it can be checked that our numerical result with radiative corrections satisfy this behavior at large $Q^2$.

\begin{table}[thb]
\begin{center}
\caption{The comparisons of  our predictions  with those  from  other different approaches for the $\rho$ meson form factors at different values of $Q^2$.
          (Note that value at $Q^2=4,5$ for light-front quark-model and $Q^2=5$ for three-point QCD sum rules are not given.) }
\label{tableff}
\vspace{0cm}
 \begin{tabular}{l |c c c} \hline \hline
$Q^2$(GeV$^2$)&~~~3~~~&~~~4~~~&~~~5~~~\\ \hline
 Form factors    &~~~$F_1$~~~$F_2$~~~$F_3$~~~& ~~~$F_1$~~~$F_2$~~~$F_3$~~~&~~~$F_1$~~~$F_2$~~~$F_3$~~~\\ \hline
     Light-front quark-model&~~~0.08~~~0.41~~~-0.10~~~& ~~~~~/~~~~~/~~~~~/~~~&~~~~~/~~~~~/~~~~~/~~~ \\
     Three-point QCD sum rules&~~~0.07~~~0.15~~~0.009~~~& ~~~0.05~~~0.10~~~0.001~~~&~~~~~/~~~~~/~~~~~/~~~ \\
     Light cone QCD sum rules&~~~0.08~~~0.17~~~-0.15~~~& ~~~0.06~~~0.16~~~-0.12~~~&~~~0.04~~~0.12~~~-0.10~~~\\
     PQCD &~~~0.07~~~0.30~~~0.20~~~& ~~~0.05~~~0.20~~~0.12~~~&~~~0.03~~~0.15~~~0.09~~~\\ \hline \hline
\end{tabular}
\end{center}
\end{table}

 The $\rho$ meson form factors have been calculated in the frameworks of different approaches, such as the three-point QCD sum rules,
the light cone sum rules, the Light-front quark-model,  the lattice QCD, and our PQCD approach.
In Table I, we present the values of the electric, magnetic and quadruple form factors predicted by different approaches at different values of $Q^2$.
At small $Q^2$, where the lattice calculation is applicable, our PQCD does not work.
So we only selected three values of $Q^2$, $Q^2=3,4,5$, at intermediate momentum transfer for QCD sum rules and large for our PQCD.
We found that our results for electric form factor $F_1$  practically identify with the prediction with three-point QCD sum rule and light cone QCD sum rule.
Then our prediction for magnetic form factor $F_2$ is 2 times larger than QCD sum rules at $Q^2=3$, while close to light-front quark-model.
But our values are getting closer to the one from QCD sum rules with the increase of $Q^2$, at $Q^2=3,4$.
The value of quadruple form factors $F_3$ are much larger than the one for other approaches, which deserve our further consideration.
Unfortunately, the PQCD does not  work properly at small $Q^2$ region,  we therefore can not determine magnetic moment of $\rho$ meson in the low-$Q^2$ region
where both the lattice calculation and the light-front quark model can work well.

\section{CONCLUSION}

In this paper we studied the $\rho \gamma^* \to \rho  $ transition process and calculated the next-to-leading-order corrections
to the $\rho$ meson electromagnetic form factors  in the framework of the PQCD factorization approach.
We firstly derive the infrared-finite NLO hard kernel for $\rho$ meson electromagnetic form factors by taking the difference of the quark
diagrams and the effective diagrams for the pion wave function,
and then calculate the NLO corrections to the form factors  $F_i(Q^2)(i=LL,LT,TL)$, and present the expressions of  the electric $F_1(Q^2)$,
the magnetic  $F_2(Q^2)$ and the quadruple $F_3(Q^2)$ form factors of $\rho$ meson in the hard scale.
We found that the NLO corrections to  $F_{LT,TL}$ are around $30\%$ of the LO contribution in the region
$Q^2 > 2 GeV^2$, where the PQCD approach is applicable.
The NLO correction to  $F_{LL}(Q^2)$ is around $20\%$ of the LO  one in the region $Q^2> 3 GeV^2$.
The NLO radiative corrections to the electric,  the magnetic, and the quadruple form factors $F_j(Q^2) (j=1,2,3)$ are sizeable in magnitude,
and  do agree with the results from the QCD sum rules.

\section{ACKNOWLEDGEMENTS}

Many thanks to Shan Cheng for valuable discussions.
This work is supported by the National Natural Science
Foundation of China under Grants  No.11847141 and 11775117,
and also by the Practice Innovation Program of Jiangsu Province under Grant No. KYCX18-1184.


\end{document}